%% file: main.tex
\documentclass[10pt,conference]{IEEEtran} 
\IEEEoverridecommandlockouts
\usepackage{acronym}
\usepackage{hyperref}
\usepackage{makecell}
\usepackage{pifont}
\usepackage{cite}
\usepackage{amsmath,amssymb,amsfonts}
\usepackage{algorithmic}
\usepackage{graphicx}
\usepackage{textcomp}
\usepackage{xcolor}
\usepackage{svg}
\usepackage{subcaption}
\usepackage{tabularx}
\usepackage{booktabs}
\usepackage{bytefield} 
\usepackage{listings}
\usepackage{tikz}
\usepackage{listings}
\usepackage{xcolor}

\definecolor{kwQuantum}{RGB}{179, 0, 0}        
\definecolor{kwScalar}{RGB}{60, 60, 60}        
\definecolor{kwVector}{RGB}{30, 90, 180}       
\definecolor{kwHalt}{RGB}{204, 102, 0}         
\definecolor{cmtColor}{RGB}{120, 120, 120}     
\definecolor{lblColor}{RGB}{0, 110, 0}         
\definecolor{bgGray}{RGB}{248, 248, 248}

\lstdefinestyle{hisepq}{
    basicstyle=\ttfamily\footnotesize,
    backgroundcolor=\color{bgGray},
    frame=single,
    framerule=0.4pt,
    rulecolor=\color{black!40},
    showstringspaces=false,
    columns=fullflexible,
    keepspaces=true,
    breaklines=false,
    numbers=left,
    numberstyle=\tiny\color{black!40},
    numbersep=5pt,
    xleftmargin=14pt,
    commentstyle=\color{cmtColor}\itshape,
    morecomment=[l]{\#},
    morecomment=[l]{//},
}

\lstdefinelanguage{HiSEPQAsm}{
    morekeywords=[1]{qv.h, qv.cx, qv.meas, qv.single, qv.resume, resume},
    morekeywords=[2]{vsetvli, vle8.v, vle32.v, vse8.v},
    morekeywords=[3]{addi, lui, jal, nop, li, la, beq, bne, j},
    keywordstyle=[1]\color{kwQuantum}\bfseries,
    keywordstyle=[2]\color{kwVector}\bfseries,
    keywordstyle=[3]\color{kwScalar},
    sensitive=true,
}

\lstdefinelanguage{QASM}{
    morekeywords=[1]{H, X, Y, Z, CX, CNOT, measure, reset, barrier},
    morekeywords=[2]{qreg, creg, OPENQASM, include, gate},
    keywordstyle=[1]\color{kwQuantum}\bfseries,
    keywordstyle=[2]\color{kwVector}\bfseries,
    sensitive=true,
}

\lstdefinelanguage{CoSimTrace}{
    morekeywords=[1]{qvsg_meas, measure_done, RESUME, MEASURE},
    morekeywords=[2]{H, CNOT, MEAS, qubit, gate, role, CTRL, TGT},
    keywordstyle=[1]\color{kwHalt}\bfseries,
    keywordstyle=[2]\color{kwQuantum}\bfseries,
    sensitive=true,
}

\usepackage{mdframed}

\newmdenv[
    topline=false,
    bottomline=false,
    rightline=false,
    linewidth=2pt,
    linecolor=black!40,
    innerleftmargin=8pt,
    innerrightmargin=0pt,
    innertopmargin=2pt,
    innerbottommargin=2pt,
    backgroundcolor=black!3
]{instrbox}

\DeclareRobustCommand*{\IEEEauthorrefmark}[1]{%
  \raisebox{0pt}[0pt][0pt]{\textsuperscript{\footnotesize\ensuremath{#1}}}}
  
\def\BibTeX{{\rm B\kern-.05em{\sc i\kern-.025em b}\kern-.08em
    T\kern-.1667em\lower.7ex\hbox{E}\kern-.125emX}}

\begin{document}

\newcommand\copyrighttext{%
  \footnotesize \textcopyright 2026 IEEE. Personal use of this material is permitted.
  Permission from IEEE must be obtained for all other uses, in any current or future
  media, including reprinting/republishing this material for advertising or promotional
  purposes, creating new collective works, for resale or redistribution to servers or
  lists, or reuse of any copyrighted component of this work in other works.}
\newcommand\copyrightnotice{%
\begin{tikzpicture}[remember picture,overlay]
\node[anchor=south,yshift=10pt] at (current page.south) {\fbox{\parbox{\dimexpr\textwidth-\fboxsep-\fboxrule\relax}{\copyrighttext}}};
\end{tikzpicture}%
}

\input{acronyms.tex}

\title{Vectorizing Quantum Control: A RISC-V Vector Extension Architecture for Scalable Qubit Systems
\thanks{
\IEEEauthorrefmark{*} Equal Contribution. }
}

\author{\IEEEauthorblockN{Xiaorang Guo\IEEEauthorrefmark{*,1}, Kun Qin\IEEEauthorrefmark{*,1,2}, Yanbin Chen\IEEEauthorrefmark{3}, Carsten Trinitis\IEEEauthorrefmark{1,2} and Martin Schulz\IEEEauthorrefmark{1}}

\IEEEauthorblockA{\small
\IEEEauthorrefmark{1}Chair of Computer Architecture and Parallel Systems, Technical University of Munich, Garching, Germany\\
\IEEEauthorrefmark{2}Chair of Computer Architecture and Operating Systems, Technical University of Munich, Heilbronn, Germany\\
\IEEEauthorrefmark{3}Chair for Formal Languages, Compiler Construction, Software Construction, Technical University of Munich, Garching, Germany\\
Email: \{xiaorang.guo, kun.qin, yanbin.chen, carsten.trinitis, martin.w.j.schulz\}@tum.de}}

\maketitle
\copyrightnotice
\begin{abstract}
The Quantum Control Processor (QCP) bridges the gap between compiler toolchains and control electronics, and is responsible for translating compiled quantum circuits into executable instructions that directly manipulate qubits and handle measurement feedback. However, existing designs rely primarily on customized instruction sets, limiting design reuse and requiring significant effort to build supporting toolchains. Furthermore, efficiently addressing qubits and scheduling operations in highly scalable scenarios remains a critical challenge. In this work, we present a vectorized quantum control approach built upon the RISC-V Vector (RVV) engine with a quantum-oriented extension. Leveraging the high parallelism of RVV, our approach can address up to 128 qubits in a single instruction. We also embed parameterized rotation information into the instruction set, enabling dynamic tuning of gate rotations in hybrid quantum-classical programs. To support mid-circuit measurements, we design a hardware-based halt-resume protocol that resumes pipeline execution within 80 $ns$ of receiving the measurement result. Comprehensive evaluation using both RISC-V toolchains and FPGA prototypes demonstrates that our design achieves up to 2.52$\times$ speedup over the baseline in program execution time, with excellent scalability.
\end{abstract}


\input{chapters/introduction}
\input{chapters/backgrounds}
\input{chapters/ISA}

\input{chapters/architecture}

\input{chapters/evaluation}
\input{chapters/relatedwork}

\input{chapters/conclusion}
\section*{Acknowledgment}
We thank Jakob Schäffeler for sharing his expertise on the RVV instruction set. We also thank Reinhard Stahn and Enrique Naranjo Bejarano from ParityQC for their helpful feedback on the design of this work. This work was funded by the German Federal Ministry of Education and Research (BMBF) under the funding program Quantum Technologies - From Basic Research to Market under contract number 13N16087, as well as from the Munich Quantum Valley (MQV), which is supported by the Bavarian State Government with funds from the Hightech Agenda Bayern.
\bibliographystyle{IEEEtran}
\bibliography{ref.bib}

\end{document}

%% file: acronyms.tex
\acrodef{CPU}[CPU]{Central Processing Unit}
\acrodef{FPGA}[FPGA]{Field-Programmable Gate Array}
\acrodef{ASIC}[ASIC]{Application-Specific Integrated Circuit}
\acrodef{GPU}[GPU]{Graphics Processing Unit}
\acrodef{QPU}[QPU]{Quantum Processing Unit}
\acrodef{NISQ}[NISQ]{Noisy Intermediate-Scale Quantum}
\acrodef{PL}[PL]{Programmable Logic}
\acrodef{PS}[PS]{Processing System}
\acrodef{DRAM}[DRAM]{Dynamic Random Access Memory}
\acrodef{DDR}[DDR]{Double Data Rate}
\acrodef{IP}[IP]{Intellectual Property}
\acrodef{MMIO}[MMIO]{Memory-Mapped I/O}
\acrodef{HLS}[HLS]{High-level Synthesis}
\acrodef{HPC}[HPC]{High Performance Computing}
\acrodef{API}[API]{Application Programming Interface}
\acrodef{QEC}[QEC]{Quantum Error Correction}
\acrodef{QCP}[QCP]{Quantum Control Processor}
\acrodef{AWG}[AWG]{Arbitrary Waveform Generator}
\acrodef{QAOA}[QAOA]{Quantum Approximate Optimization Algorithm}
\acrodef{VQE}[VQE]{Variational Quantum Eigensolver}
\acrodef{ISA}[ISA]{Instruction Set Architecture}
\acrodef{SIMD}[SIMD]{Single Instruction, Multiple Data}
\acrodef{LMUL}[LMUL]{Length Multiplier}
\acrodef{ADC}[ADC]{Analog-to-Digital Converter}
\acrodef{MCM}[MCM]{Mid-Circuit Measurement}
\acrodef{FTQC}[FTQC]{Fault-Tolerant Quantum Computing}

%% file: chapters/introduction.tex
\section{Introduction}
Quantum computing is moving from theoretical promise to physical realization, with leading modalities such as superconducting qubits~\cite{huang2020superconducting}, trapped 
ions~\cite{bruzewicz2019trapped}, and neutral atoms~\cite{wintersperger2023neutral} demonstrating qubit counts ranging from tens to thousands in recent years~\cite{bluvstein2026fault,abughanem2025ibm}. Yet achieving practical quantum advantage requires more than physical-level improvements: it demands coordinating qubit manipulation at scale. This challenge falls primarily on the classical control hardware that drives the \ac{QPU}. As qubit counts grow beyond a thousand toward \ac{QEC} applications, the classical control stack emerges as a critical scalability bottleneck that limits further system expansion.

A modern quantum computing stack is composed of heterogeneous layers: from high-level algorithms and compilers down to system control software, cryogenic electronics, and the physical quantum modality~\cite{microsoft_quantum_stack}. The classical control bottleneck primarily occurs at the \textit{system control software} layer, where two demands grow simultaneously as systems scale. First, as qubit counts increase, the control unit must issue gate operations to more qubits in parallel, placing ever-greater scheduling pressure on the classical hardware. Second, the rise of hybrid quantum-classical algorithms, such as the \ac{VQE}~\cite{peruzzo2014variational} and the \ac{QAOA}~\cite{farhi2014quantum}, requires low-latency, feedback-driven execution, where measurement outcomes from the \ac{QPU} must be processed by classical logic before the next operation is issued. In practice, end-to-end profiling of such workloads shows that total execution time is dominated not by quantum gate operations themselves, but by the round-trip communication between the \ac{QPU} and the classical host~\cite{Qtenon}, making the classical-quantum interface the primary performance bottleneck. Taken together, these demands call for a dedicated \textit{\ac{QCP}} that can serve as a programmable, scalable bridge between classical computation and physical qubit manipulation.

Several \acp{QCP} have been proposed to fill this gap, ranging from \ac{FPGA}-based control platforms~\cite{stefanazzi2022qick,liu2025risc, xu2023qubic, xu2021qubic, GuoHiSEP} to dedicated processor architectures~\cite{fu2019eqasm,fu2018quma,Qtenon}. 
Despite their contributions, these efforts suffer from one or more of the following fundamental limitations. First, they mostly rely on custom \acp{ISA} that require a specific compiler to translate quantum circuits into dedicated instructions, making their practical use difficult and compiler development expensive. Second, their parallel gate-scheduling and dispatch mechanisms are either absent or tightly coupled to a specific physical modality, providing no clear scaling path toward the qubit counts required for \ac{QEC}. Third, without a standardized \ac{ISA} substrate, these designs lack access to mature compiler infrastructure, formal verification frameworks, and the broader open-source hardware ecosystem, which are prerequisites for long-term adoption. 

These challenges point toward a natural candidate: the RISC-V \ac{ISA}~\cite{rvv_spec}. As an open and standard \ac{ISA}, RISC-V supports domain-specific extensions through 
reserved opcode space without breaking binary compatibility, and its vector extension (RVV)~\cite{rvv_spec} already provides the configurable, length-agnostic \ac{SIMD} semantics that quantum gate dispatch demands, enabling parallel targeting of multiple qubits within a single instruction. Most importantly, any RISC-V extension directly inherits a mature ecosystem, including LLVM compiler backends, open-source processor cores, and standard co-processor interfaces~\cite{xif_spec}, thereby eliminating the toolchain burden that has hindered prior \ac{QCP} designs.

Therefore, in this paper, we propose HiSEP-Q 2.0: \textit{A RISC-V Vector Extension Architecture for Scalable Qubit Systems}. We integrate the qubit-targeting mechanism of HiSEP-Q~\cite {GuoHiSEP} into the RVV extension framework, where we encode qubit indices as packed 8-bit vector elements (configurable) and use the vector grouping mechanism to 
address up to 128 qubits in a single instruction, achieving high code density and large gate-level parallelism. Additionally, we extend the \ac{ISA} with dedicated rotation instructions that carry per-qubit angle operands, enabling seamless support for the parameterized gates required by hybrid quantum-classical algorithms. At the microarchitecture level, we build upon open-source RISC-V projects~\cite{ibexDocs, platzer2021vicuna} and contribute a measurement-based halt-resume protocol that tightly coordinates the control core and the vector pipeline across qubit measurement boundaries. The full design is implemented at the Register-Transfer Level (RTL), and subsequently verified and evaluated using both RISC-V and FPGA toolchains. 
Overall, our contributions are as follows:
\begin{itemize}
    \item We propose a quantum extension to the RISC-V Vector \ac{ISA} that scales parallel gate dispatch from 8 to 128 qubits per instruction via \ac{LMUL} grouping, and embeds gate rotation operands to support parameterized gates in hybrid programs.

    \item We design a halt-resume protocol for \ac{MCM} support that re-streams the pipeline within 80 $ns$, enabling measurement-conditional branching for adaptive quantum algorithms well within typical qubit coherence windows.

    \item We deliver a basic automatic Open Quantum Assembly Language (OpenQASM)-to-binary compilation flow that can lower quantum circuits onto our \ac{QCP} without manual intervention.

    \item We provide a complete RTL implementation on the Xilinx ZCU216 FPGA and evaluate it across 12 Munich Quantum Toolkit (MQT) Bench circuits, achieving a geometric-mean speedup of $1.32\times$ (peak $2.52\times$) over HiSEP-Q~1.0, while consuming less than 3.5\% of the device resources.

\end{itemize}



%% file: chapters/backgrounds.tex
\section{BACKGROUND}

\subsection{Quantum Control Processor (QCP)}

\acp{QCP} are increasingly recognized as the critical system-level interface bridging high-level programming environments and \acp{QPU} backends\cite{britt2017isaqpu, fu2019eqasm,Qtenon,HISQ,GuoHiSEP}. Modern advanced algorithms, particularly hybrid classical-quantum routines, typically deviate from a linear execution of gate sequences. Instead, they require dynamic control flow driven by real-time feedback. The need for low-latency, efficient feedback handling has motivated the emergence of dedicated, hardware-implemented control units that can be positioned as close to the \acp{QPU} as possible.

From a functionality point of view, a \ac{QCP} translates compiled programs into precisely timed gate sequences, which are then sent to ~\acp{AWG} for physical manipulation. Each gate corresponds to a specific pulse shape and duration, and the AWG must be triggered at an exact scheduled time to drive the target qubit correctly. For multi-qubit gates such as the controlled-NOT (CNOT), the control and target qubits must receive their respective pulses simultaneously — any timing skew beyond the pulse envelope degrades gate fidelity. This cycle-accurate firing requirement places a hard real-time constraint on the \ac{QCP}. In other words, it must guarantee that all qubits are issued the corresponding gates at scheduled timestamps.

A further challenge arises from \ac{MCM}, a primitive essential to \ac{QEC} and adaptive algorithms. Unlike terminal measurements, mid-circuit measurement interrupts the gate sequence at an intermediate point, which means the \ac{QCP} issues a measurement pulse, suspends further gate scheduling, and waits for \acp{ADC} to return a classical readout result. This readout latency is protocol-dependent and not known at compile time, introducing an unbounded pause into an otherwise deterministic instruction stream. Upon receiving the measurement outcome, the \ac{QCP} must resume execution and, in general, select subsequent gates conditionally based on the classical result. This feedback loop, including gate scheduling, measurement, classical decision, and conditional resumption, defines the fundamental control challenge that a \ac{QCP} must address in hardware.

\subsection{RISC-V and Vector Extension}

RISC-V is a suitable foundation because it provides an open ISA with a modular extension structure \cite{rvv_spec,schlaegl2024rvvvp}. This property is important for quantum control processors, since they still require ordinary instruction-level mechanisms such as control flow, arithmetic, and memory access around quantum-specific operations \cite{britt2017isaqpu,GuoHiSEP}. An extension-based ISA is therefore a cost-efficient way to integrate classical control behavior and domain-specific quantum operations within a unified framework \cite{britt2017isaqpu,rvv_spec}. This motivation further suggests a SIMD-oriented architectural model, as a quantum control processor must often address tens to hundreds of qubits with a single instruction to realize efficient data and control parallelism.

Therefore, RISC-V Vector Extension (RVV) becomes a natural choice that meets the QCP requirements. It provides a dedicated architectural mechanism for SIMD execution in the RISC-V ecosystem. By enabling a single instruction to process multiple elements in parallel, RVV offers a natural solution for applications that require structured parallel quantum operations. RVV provides flexibility at multiple levels. It first removes the dependence on a fixed hardware vector width by determining the active vector length at runtime. Based on this length-agnostic feature, software can further configure vector organization through parameters, as shown in Table~\ref{tab:rvv_terms}. RVV also supports masked execution and well-defined inactive-element semantics, so quantum operations can be applied only to selected elements. This selective execution model is useful for quantum control, where operations often target specific subsets rather than all indexed elements uniformly \cite{rvv_spec,llvm_rvv,fu2019eqasm,GuoHiSEP}.

\begin{center}
\begin{bytefield}[bitwidth=0.75em,endianness=big]{32}
  \bitheader{31,25,24,20,19,15,14,12,11,7,6,0} \\
  \bitbox{7}{\scriptsize funct7}
  & \bitbox{5}{\scriptsize operand 2}
  & \bitbox{5}{\scriptsize operand 1}
  & \bitbox{3}{\scriptsize funct3}
  & \bitbox{5}{\scriptsize result reg}
  & \bitbox{7}{\scriptsize Opcode/OP-V}
\end{bytefield}
\end{center}

As shown in the bitmap above, RISC-V provides an extensible ISA, while RVV shapes a scalable data-parallel model for manipulating indexed qubit sets and their associated parameters. The RVV properties are appealing for a \ac{QCP} design because future implementations may scale the available vector resources without requiring a different programming model \cite{schlaegl2024rvvvp}. This motivates the vector-oriented design adopted in this work.

\begin{table}[h]
\centering
\caption{RVV Terms}
\label{tab:rvv_terms}
\small
\begin{tabularx}{\columnwidth}{l >{\raggedright\arraybackslash}X}
\toprule
\textbf{Term}  & \textbf{Explanation} \\
\hline
\texttt{VLEN}  & physical vector register bit-width \\
\texttt{VL} & Active number of elements \\
\texttt{AVL}  & Number of elements requested by software \\
\texttt{VLMAX} & Maximum number of elements under \texttt{SEW} and \texttt{LMUL} \\
\texttt{SEW} & Selected element width in bits, 8/16/32 \\
\texttt{LMUL} & Combines multiple vector registers into one group \\
\bottomrule
\end{tabularx}
\end{table}

Our architecture builds on two existing open-source components: the Ibex RV32IMC core and the Vicuna vector coprocessor. Ibex acts as the scalar host processor and provides the general-purpose \textsc{RISC-V} execution environment for control and system software, while Vicuna provides the vector-engine substrate reused for offloaded execution, vector-register handling, and pipeline-based processing. Vicuna is particularly attractive for this work
because it was designed as a time-predictable vector coprocessor, a property that is highly valuable for quantum event generation and measurement-aware execution control. In general, these two components provide the classical foundation on top of which the proposed quantum extensions are built.

%% file: chapters/ISA.tex
\section{Instruction Set Architecture}
\label{sec:isa}

The quantum \ac{ISA} serves as the interface between quantum programs and their hardware controllers. To support the execution of quantum circuits, \ac{ISA} typically includes three categories of instructions~\cite{GuoHiSEP,fu2019eqasm}: classical instructions for control flow and register updates, timing directives for gate scheduling, and quantum instructions for qubit selection and gate specification. This section introduces how we design our \ac{ISA} to be compatible with the RISC-V ecosystem while supporting all three categories of instructions.

\subsection{Extension Overview}

The HiSEP-Q~2.0 \ac{ISA} extends the RISC-V Vector (RVV) specification
rather than defining a standalone quantum instruction set. By occupying
reserved encoding space within the existing OP-V major opcode
(\texttt{1010111}), the extension remains binary-compatible with the
standard RVV toolchain. Classical control flow, memory management,
branch logic and timing control are handled entirely by standard RISC-V instructions on
the Ibex scalar core~\cite{ibexDocs}, while gate dispatch and
measurement are offloaded to the vector co-processor (qvproc) via the
CORE-V XIF interface~\cite{xif_spec}. The quantum extension introduces four new instructions, summarised in Table~\ref{tab:isa_summary}, organised
around three operational classes: single-qubit gates, two-qubit
entangling gates, and parameterised rotation gates. The \texttt{funct3}
field distinguishes the class; the upper \texttt{funct7}/GateID bits
identify the specific gate within each class.

\begin{table}[h]
\centering
\caption{HiSEP-Q~2.0 Quantum Vector Instructions}
\label{tab:isa_summary}
\small
\begin{tabularx}{\columnwidth}{l l >{\raggedright\arraybackslash}X}
\toprule
\textbf{Instr.} & \textbf{funct3} & \textbf{Purpose} \\
\midrule
\texttt{QV.SINGLE} & \texttt{000}
    & Single-qubit gates on a vector of qubit indices\\
\texttt{QV.PAIR}   & \texttt{001}
    & Two-qubit gates on element-wise control--target pairs\\
\texttt{QV.ROT.G}  & \texttt{010}
    & One global fixed-point angle  broadcast to all target qubits \\
\texttt{QV.ROT.V}  & \texttt{011}
    & Per-qubit variable rotation \\
\bottomrule
\end{tabularx}
\end{table}

\subsection{Qubit Addressing}



RVV provides highly scalable qubit-addressing capacity. In the current design, physical qubit indices are stored as 8-bit unsigned integers (tracking up to 256 qubits) packed into standard vector registers (\texttt{v0}--\texttt{v31}). Thus, the extended \ac{ISA} can address up to 128 qubits per instruction, leading to large parallelization opportunities to achieve \ac{SIMD} property. The width of the indices can be configured freely via \texttt{SEW}, depending on the size of quantum processors. Each vector register of width \texttt{VLEN} bits is subdivided into $\texttt{VLEN}/\texttt{SEW}$ index slots, each holding one qubit index. The number of qubits addressed per instruction is determined by the standard \texttt{vsetvli} configuration (\texttt{SEW}, \texttt{LMUL}):
\begin{equation}
N_{\text{qubits}} = \text{VL} = \min\!\left(\text{AVL},\;
\frac{\text{VLEN} \times \text{LMUL}}{\text{SEW}}\right)
\label{eq:vl}
\end{equation}
where AVL is set to \texttt{VLMAX} in practice to utilize the full register capacity.

With \texttt{VLEN=128} and \texttt{SEW=8}, the maximum parallelism
ranges from 8 qubits (\texttt{LMUL=mf2}) to 128 qubits
(\texttt{LMUL=m8}), as shown in Table~\ref{tab:lmul}. This scalability
requires no architectural changes: a single \texttt{vsetvli} instruction
selects the desired configuration. Furthermore, the instruction encoding also carries a \texttt{Blk\_imm} field (5 bit) that is forwarded to the quantum backend as scheduling metadata for timing control.

\begin{table}[h]
\centering
\caption{Qubit parallelism vs.\ LMUL (\texttt{VLEN=128}, \texttt{SEW=8})}
\label{tab:lmul}
\small
\begin{tabular}{ccc}
\toprule
\textbf{LMUL} & \textbf{Physical register group} & \textbf{Max qubits per instr.} \\
\midrule
\texttt{mf2} & 0.5 registers &   8 \\
\texttt{m1}  & 1 register    &  16 \\
\texttt{m2}  & 2 registers   &  32 \\
\texttt{m4}  & 4 registers   &  64 \\
\texttt{m8}  & 8 registers   & 128 \\
\bottomrule
\end{tabular}
\end{table}


\subsection{Quantum Instruction Encoding and Semantics}
This subsection details the encoding and semantics of the four quantum vector instructions introduced above.

\subsubsection{QV.SINGLE: Single-Qubit Gate}

\texttt{QV.SINGLE} applies one single-qubit gate to every qubit index stored 
in \texttt{vs1}. The gate type is selected by the 7-bit \texttt{GateID}
field in bits~[31:25]. 

\begin{center}
\begin{bytefield}[bitwidth=0.75em,endianness=big]{32}
  \bitheader{31,25,24,20,19,15,14,12,11,7,6,0} \\
  \bitbox{7}{\scriptsize GateID}
  & \bitbox{5}{\scriptsize rs2}
  & \bitbox{5}{\scriptsize vs1}
  & \bitbox{3}{\scriptsize 000}
  & \bitbox{5}{\scriptsize Blk\_imm}
  & \bitbox{7}{\scriptsize 1010111}
\end{bytefield}
\end{center}

\subsubsection{QV.PAIR: Two-Qubit Gate}

\texttt{QV.PAIR} applies a two-qubit gate to arbitrary element-wise pairs drawn
from two vector registers: \texttt{vs2} carries the control qubit indices
and \texttt{vs1} carries the target qubit indices. The $i$-th element of
\texttt{vs2} is paired with the $i$-th element of \texttt{vs1}, so both
registers must be loaded with the same active vector length.

\begin{center}
\begin{bytefield}[bitwidth=0.75em,endianness=big]{32}
  \bitheader{31,25,24,20,19,15,14,12,11,7,6,0} \\
  \bitbox{7}{\scriptsize GateID}
  & \bitbox{5}{\scriptsize vs2(src)}
  & \bitbox{5}{\scriptsize vs1(tgt)}
  & \bitbox{3}{\scriptsize 001}
  & \bitbox{5}{\scriptsize Blk\_imm}
  & \bitbox{7}{\scriptsize 1010111}
\end{bytefield}
\end{center}

\subsubsection{QV.ROT.G: Global Rotation}

\texttt{QV.ROT.G} applies the same rotation angle to every target qubit
in \texttt{vs1}. The angle is a 32-bit fixed-point value held in the
integer scalar register \texttt{rs2}, forwarded unchanged to the backend.
The fixed-point encoding (e.g., a full-scale mapping to $2\pi$) is a
system-level convention; the hardware treats it as an opaque 32-bit word.

\begin{center}
\begin{bytefield}[bitwidth=0.75em,endianness=big]{32}
  \bitheader{31,25,24,20,19,15,14,12,11,7,6,0} \\
  \bitbox{7}{\scriptsize Res}
  & \bitbox{5}{\scriptsize rs2(ang)}
  & \bitbox{5}{\scriptsize vs1(tgt)}
  & \bitbox{3}{\scriptsize 010}
  & \bitbox{5}{\scriptsize Blk\_imm}
  & \bitbox{7}{\scriptsize 1010111}
\end{bytefield}
\end{center}

\subsubsection{QV.ROT.V: Variable Rotation}

\texttt{QV.ROT.V} extends \texttt{QV.ROT.G} to per-qubit rotation angles, which means each qubit can have different rotations.
Target qubit indices are read from \texttt{vs1} at \texttt{SEW=8}, while
the corresponding rotation angles are read from \texttt{vs2} at
\texttt{SEW=32}. The $i$-th 32-bit angle in \texttt{vs2} is applied to
the $i$-th 8-bit qubit index in \texttt{vs1}.

\begin{center}
\begin{bytefield}[bitwidth=0.75em,endianness=big]{32}
  \bitheader{31,25,24,20,19,15,14,12,11,7,6,0} \\
  \bitbox{7}{\scriptsize Res}
  & \bitbox{5}{\scriptsize vs2(ang)}
  & \bitbox{5}{\scriptsize vs1(tgt)}
  & \bitbox{3}{\scriptsize 011}
  & \bitbox{5}{\scriptsize Blk\_imm}
  & \bitbox{7}{\scriptsize 1010111}
\end{bytefield}
\end{center}

Because \texttt{vs2} elements are 4$\times$ wider than \texttt{vs1}
elements, the two operands occupy register groups of different sizes. The hardware implicitly scales the \texttt{vs2} group by a factor of
four relative to the \texttt{vs1} configuration:

\begin{equation}
\text{LMUL}_{\texttt{vs2}} = 4 \times \text{LMUL}_{\texttt{vs1}}
\label{eq:lmul_scale}
\end{equation}

This mixed-precision design avoids the need for two decoupled execution
pipelines. Instead of treating the index and angle operands as
independent data streams requiring separate hardware coordination,
the alignment between the two register groups is derived statically
at decode time from the existing \texttt{vsetvli} configuration.
This reuse of the native LMUL mechanism adds no runtime overhead and
requires no changes to the standard RVV decode logic beyond the
implicit group-size scaling rule above.

However, as a tradeoff, this mechanism also limits the legal \texttt{vs1} configurations to \texttt{mf2},
\texttt{m1}, and \texttt{m2}. In other words, \texttt{m4} and \texttt{m8} would require
a \texttt{vs2} group exceeding the maximum supported (\texttt{m8}) and
are therefore rejected as illegal by the hardware. The legal configurations and resulting element counts are summarised in Table~\ref{tab:qrv_lmul}.

\begin{table}[h]
\centering
\caption{\texttt{QV.ROT.V} legal LMUL configurations}
\label{tab:qrv_lmul}
\small
\begin{tabular}{cccc}
\toprule
\textbf{vs1} & \textbf{vs2 (implicit)} & \textbf{Elements} & \textbf{Status} \\
\midrule
\texttt{e8,\,mf2} & \texttt{e32,\,m2}  &  8 & Legal   \\
\texttt{e8,\,m1}  & \texttt{e32,\,m4}  & 16 & Legal   \\
\texttt{e8,\,m2}  & \texttt{e32,\,m8}  & 32 & Legal   \\
\texttt{e8,\,m4}  & \texttt{e32,\,m16} & 64 & Illegal \\
\texttt{e8,\,m8}  & \texttt{e32,\,m32} & 128 & Illegal \\
\bottomrule
\end{tabular}
\end{table}

The compiler needs to  additionally verify that the expanded \texttt{vs2}
register group does not overlap with other live vector register
allocations, and must emit separate load instructions for the two operands: \texttt{vle8.v} for \texttt{vs1} and \texttt{vle32.v} for \texttt{vs2}.

Additionally, as seen in the aforementioned instructions, each quantum vector instruction
carries a 5-bit immediate named \texttt{Blk\_imm}.
This field encodes
a programmer-defined delay interval between consecutive gate operations,
allowing the backend to insert the necessary wait time between instructions for timing control.


%% file: chapters/architecture.tex
\section{Architecture}
\subsection{System Overview}
\begin{figure}[b]
    \centering
    \includegraphics[width=0.8\linewidth]{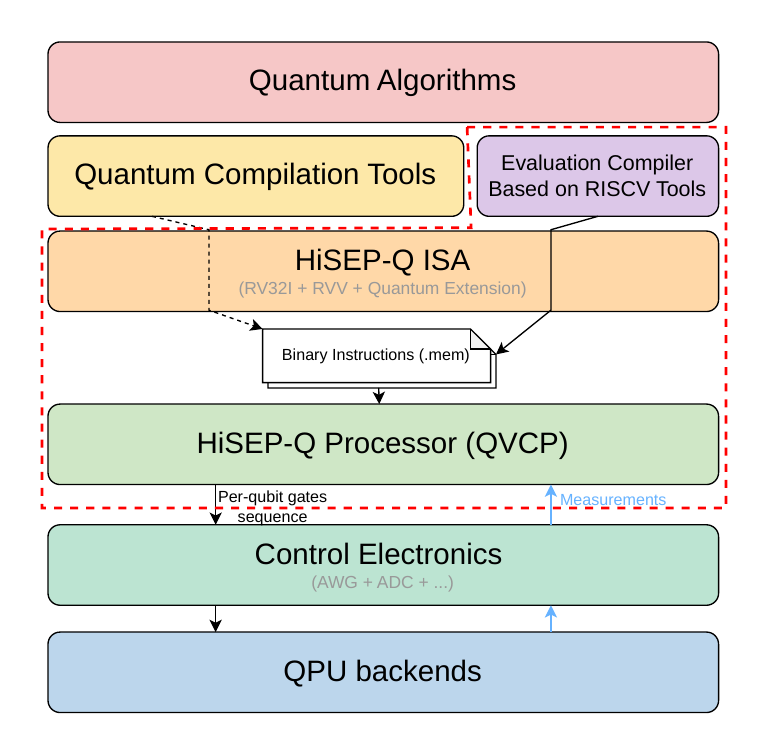}
    \caption{HiSEP-Q~2.0 in the quantum software stack. The dashed boundary marks the components contributed by this work.}
    \label{fig:stack}
\end{figure}

\begin{figure*}[t]
    \centering
        \includegraphics[width=0.925\linewidth]{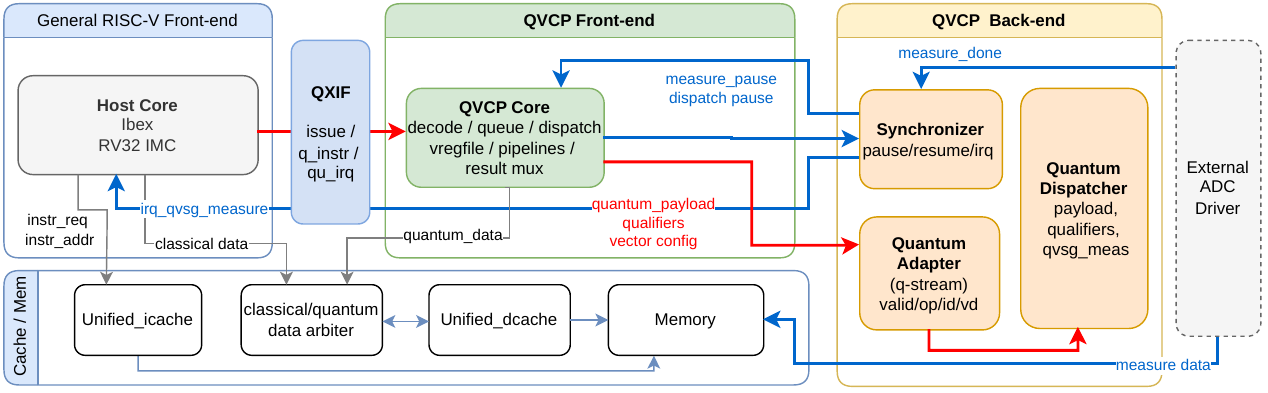}
    \caption{Overall architecture of the Ibex-coupled quantum vector engine and its execution back end.}
    \label{fig:general_archi}
\end{figure*}

Figure~\ref{fig:stack} illustrates how HiSEP-Q~2.0 fits into the broader quantum software stack. The typical stack spans from high-level quantum algorithms at the top, through compilation tools that lower circuits to executable binaries, down to the control electronics (AWG, ADC) that drive the physical \ac{QPU} backend. The overall contributions of this work, highlighted by the red dashed boundary, comprise three
tightly coupled components: (1)~the quantum \ac{ISA}, which extends RV32I and the RVV specification with quantum-specific semantics; (2)~the QVCP, our microarchitectural realization of
the \ac{ISA} as a vector control processor; and (3)~an evaluation compiler built on the existing RISC-V toolchain that translates quantum circuits into binaries
executable by the QVCP. Note that advanced compilation passes and optimizations are beyond the scope of this work. HiSEP-Q~2.0 is open-source and available via GitHub\footnote{https://github.com/caps-tum/HiSEP-Q-2.0}.

\subsection{Microarchitecture}
    
At the microarchitecture level, the proposed processor forms a closed execution loop spanning the host core, the quantum-capable vector engine, the shared memory hierarchy, and the external quantum readout environment. As shown in Figure~\ref{fig:general_archi}, the host-side execution domain is the \textbf{General RISC-V Front-end}, while the main execution domain for vector and quantum instructions is the QVCP Core inside the \textbf{QVCP Front-end}. The two domains are coupled by the QXIF interface, which transports offloaded instructions and related metadata from the host and returns interrupt requests in the opposite direction.

This coupling allows the host processor and the quantum-vector front-end to collaborate without collapsing into a single monolithic execution pipeline. The host continues to provide scalar control, ordinary instruction sequencing, and software-visible machine state, whereas the QVCP Core performs the vector-oriented work: instruction decode, queue management, dispatch, vector-register access, pipeline execution, and result multiplexing. Quantum instructions thus appear in software as part of a single program stream, but their detailed handling is delegated to the QVCP Core, which extends the vector engine with quantum semantics.

The cache and memory subsystem provides the shared transport layer that sustains both execution domains. Instruction fetch remains on the host side through the instruction-cache path, while classical and quantum data requests converge through a data arbiter into the unified data-cache path. These requests are then forwarded to the external memory interface. This shared organization is important because it shows that the quantum overlay does not introduce an independent memory system; instead, it builds directly on the existing cache- and arbiter-based infrastructure of the
host-vector platform.


The quantum extension is implemented as a structured overlay on top of the existing vector engine. The essential idea is that quantum instructions are processed and scheduled in a manner similar to that of ordinary vector instructions, but their semantics are captured and exported through a dedicated quantum sideband rather than solely as vector-register updates. This organization aligns with prior quantum microarchitecture proposals that separate instruction transport from quantum back-end actuation and readout \cite{fu2018quma,fu2019eqasm,britt2017hpcqpu}.

\begin{figure}[htbp]
    \centering
    \includegraphics[width=0.8\columnwidth]{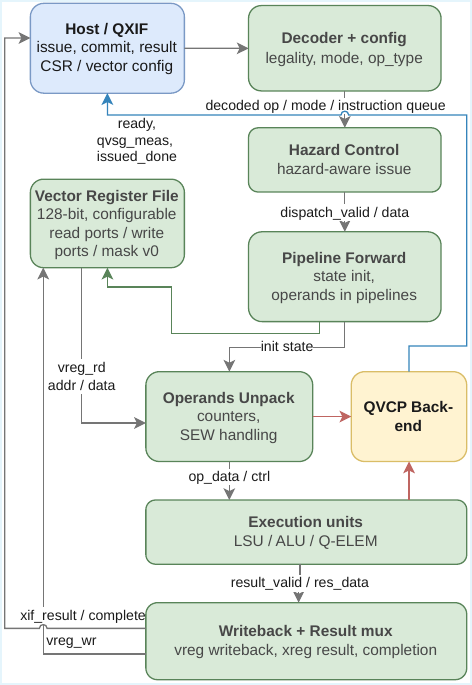}
    \caption{Main blocks of \texttt{QVCP Core}. Quantum instructions use a dedicated sideband export path to expose the quantum elements to the QVCP Back-end.}
    \label{fig:qvcp-core}
\end{figure}

\textbf{QVCP Core} inside \textbf{QVCP Front-end} handles the main part of the quantum processing, as shown in Figure \ref{fig:qvcp-core}. A quantum instruction first arrives from the host side through the \texttt{QXIF} offload interface. The \emph{Decoder} block then determines the operation class, checks legality under the current vector configuration, and assigns operand roles. The \emph{Dispatcher} selects the decoded instruction when the required execution resources and register-file dependencies permit forward progress. Up to this point, quantum instructions behave like ordinary offloaded vector work: they reuse the same structural mechanisms for admission control and hazard management.

Once dispatched, the instruction is translated into pipeline-local state by the \emph{Pipeline wrapper}. This stage fixes the operand mapping and forwards the relevant metadata into the execution path. The subsequent \emph{Pipeline + unpack} block is responsible for operand movement, counter progression, and element-width handling while reading the required data from the \emph{Vector register file}, which remains the common storage substrate for both conventional vector operands and the qubit-index or parameter streams used by the quantum extension.

Execution then proceeds in the \emph{Execution units} block. Ordinary vector operations are handled by units such as the LSU and ALU, whereas quantum instructions are steered to the quantum-capable element path, denoted here as \emph{Q-ELEM}, where the vector engine produces the quantum-visible per-element semantics. The ordinary architectural result path continues through \emph{Writeback + result mux}, which merges vector-register writeback, scalar result return, and completion signaling back toward the host. It exports a dedicated quantum sideband in parallel through the \emph{QVCP Back-end} block. This path carries the externally visible quantum payload fields, including \texttt{valid}, \texttt{op\_type}, \texttt{id}, and quantum payloads, together with the associated qualifiers and measurement-related control signals. Hence, the quantum-extended vector engine emits a structured quantum event stream for external consumption.

The \textbf{QVCP Back-end} is composed of Synchronizer and Quantum Adapter. The Synchronizer manages the control plane. This module coordinates pauses, resumptions, and interrupt-related behavior for the new instructions according to the different semantics of quantum operations. For example, when a measurement instruction is recognized, the Synchronizer asserts the appropriate control signals, allowing the active quantum stream to drain, emits the corresponding issue-done notification, and then waits for the external readout completion signal \texttt{measure\_done} from the ADC. Hence, the Synchronizer does not replace the data path execution of the measurement instruction; it wraps that execution in a hardware-managed control flow.

The \emph{Quantum Adapter} corresponds to the export layer. It gathers the data streams generated by the quantum vector engine, including the payload fields, associated qualifiers, and consumer-ready semantics for quantum electronics. In this way, the adapter does not compute quantum operations itself; rather, it translates internal per-lane activity into an externally consumable quantum event stream for further processing. 

\emph{Quantum Dispatcher} sits behind this interface and turns the \emph{Quantum Adapter}'s event stream into per-qubit, time-scheduled firing pulses. Each incoming event is decoded into a qubit identifier, a gate identifier, and a scheduled dispatch time (current global
counter $+$ \texttt{block\_imm}); the event is then enqueued into
the destination qubit's \texttt{timed\_fifo}, and a per-qubit time-controller releases it when the global counter reaches the
scheduled time.

\subsection{Distinctive Features of the Quantum Overlay}
First, the proposed quantum overlay supports mixed-precision operand handling within a single quantum instruction. This capability is exemplified by \texttt{QV.ROT.V}, where the qubit-index stream is carried as 8-bit elements while the associated rotation-angle stream is carried as 32-bit elements. In standard RVV execution, a single instruction is typically governed by one active \texttt{SEW/LMUL} configuration, although the ISA does support special cases of asymmetric operand-width. However, this must be handled by configuring the vector layout in multiple instructions, which reduces program efficiency. The present design extends this idea further for quantum operations by allowing two semantically paired operand streams within only one instruction to remain at different precisions throughout execution. The narrow path preserves a compact representation for qubit indices, while the wider path preserves the numerical precision required for rotation parameters. This behavior is implemented directly in the microarchitecture through operand-role assignment, per-operand width handling, unpack logic, and element-wise execution, so that one quantum instruction can naturally combine compact index transport with higher-precision parameter delivery.

Second, the quantum overlay has dedicated control treatment of measurement. Rather than requiring software to approximate the measurement boundary through inserted \texttt{WAIT}, \texttt{NOP}, or other timing-padding instructions, the design resolves the measurement protocol directly at the microarchitectural level. Once a measurement instruction is recognized, the hardware asserts the corresponding control state, permits the active quantum stream to drain to a defined completion point, emits a measurement-issued notification, and then waits for the external readout completion signal before releasing execution of the next instruction for either scalar or quantum-vector processing. This choice is important because the safe halt and resume boundaries depend on in-flight pipeline state and on the arrival time of the external measurement response, neither of which can be reliably encoded as a fixed compile-time software delay. By handling this protocol in hardware, the overlay avoids software-visible delay management and maintains a precise execution boundary for measurement-aware control.

%% file: chapters/evaluation.tex
\section{Evaluation}
In this section, we first introduce the experimental setup, including the evaluation toolchain, the experimental hardware, and the compilation tools. We then present the metrics used for performance evaluation and verification. 
\subsection{Experimental Setup}
HiSEP-Q~2.0 is implemented in Verilog and SystemVerilog and synthesized for the Xilinx Zynq UltraScale+ RFSoC ZCU216 evaluation board~\cite{AMDrfsoc}, which is a typical FPGA board for quantum control systems. The processor is clocked at $100~\text{MHz}$. 

\textbf{Evaluation Compiler:}
All benchmark circuits are specified in OpenQASM 3.0 and compiled to executable HiSEP-Q~2.0 binaries using a custom two-stage toolchain. In the first stage, a QASM front end parses the circuit, maps logical qubits to contiguous 8-bit physical indices, and lowers supported operations into vectorized assembly, while emitting the required scalar and RVV setup instructions. In the second stage, a lightweight back end translates the resulting assembly into 32-bit machine words for both the standard RV32I/RVV subset and the custom HiSEP-Q quantum instructions. To independently validate the standard portion of the binary, every program is also passed through the upstream RISC-V GNU toolchain as a compatibility check.

\subsection{End-to-End Workflow Verification}
\label{sec:workflow}

We verify the HiSEP-Q~2.0 end-to-end workflow, which traces from a high-level quantum program to the per-qubit dispatch events emitted at the hardware
boundary, on a representative 8-pair Bell-state circuit. Larger
benchmarks from MQT Bench~\cite{quetschlich2023mqt} are evaluated in the following section.

\subsubsection{\textbf{Quantum Circuit}}
The Bell-state program (Listing~\ref{lst:qasm}) prepares eight
entangled pairs $(q[2i], q[2i+1])$, $i\in\{0,\dots,7\}$. It exercises three primitives in a single source: parallel single-qubit gates (Hadamard), parallel two-qubit gates (CNOT), and measurement with classical-side halt-resume synchronization.

\begin{figure}[htbp]
\vspace{-8pt}
\centering
\begin{lstlisting}[style=hisepq, language=QASM,
    caption={QASM 3.0 Bell-state source (8 pairs).},
    label={lst:qasm}]
OPENQASM 3.0;
include "stdgates.inc";
qubit[16] q;  bit[8] c;
for int i in [0:7] {
    h q[2*i];
    cx q[2*i], q[2*i+1];
    measure q[2*i] -> c[i];
}
\end{lstlisting}
\vspace{-8pt}
\end{figure}

\subsubsection{\textbf{Assembly and Binary Encoding}}
Listing~\ref{lst:program} shows the program in two equivalent views.
Because each RVV instruction operates element-wise on a
qubit-index vector, the eight Hadamards collapse into a single
\texttt{qv.h}; CNOT and MEASURE follow the same pattern. The program fits in 14 words in total: a one-time setup (scalar tags, \texttt{vsetvli} with $\text{VL}=\text{vlmax}=8$, two \texttt{vle8.v} loads) followed by four RVV-format quantum instructions, each carrying its gate identifier in bits[31:25] (\texttt{0x64}=H, \texttt{0x66}=CNOT, \texttt{0x68}=MEASURE, \texttt{0x78}=resume marker). For comparison, the same circuit in the QIR representation~\cite{Amr:hpc} expands to 24 separate \texttt{\_\_quantum\_\_qis\_\*} calls (one per gate or measurement), making the vectorized HiSEP-Q~2.0 encoding roughly $6\times$ denser at the gate level (24 calls vs. 4 RVV instructions).

\begin{figure}[htbp]
\vspace{-10pt}
\centering
\begin{lstlisting}[style=hisepq, escapeinside={(*@}{@*)},
    caption={Bell-state preparation program},
    label={lst:program}]
Addr  Binary       Assembly
----  ----------   ---------------------------
0x00  06600313     addi    x6, x0, 0x66
# resume tag (carried by qv.resume)
0x04  05500393     addi    x7, x0, 0x55
# gate tag (carried by qv.h / qv.meas)
0x08  00800293     addi    x5, x0, 8
# AVL = 8 (= vlmax for mf2)
0x0C  0C72F057     vsetvli x0, x5, e8, mf2, ta, ma
0x10  00000013     nop
0x14  00001537     lui     a0, 0x1
0x18  02050087     vle8.v  v1, (a0)
# v1 <- 8 control qubit indices
0x1C  00850593     addi    a1, a0, 8
0x20  02058107     vle8.v  v2, (a1)
# v2 <- 8 target qubit indices
0x24  (*@\textcolor{kwQuantum}{\textbf{C8708657}}@*)     (*@\textcolor{kwQuantum}{\textbf{qv.h}}@*)     v3, v1, x7, 12
# gate ID 0x64 -> 8 parallel H gates
0x28  (*@\textcolor{kwQuantum}{\textbf{CC208657}}@*)     (*@\textcolor{kwQuantum}{\textbf{qv.cx}}@*)    v3, v1, v2, 12
# gate ID 0x66 -> 8 parallel CNOTs
0x2C  (*@\textcolor{kwQuantum}{\textbf{D0708657}}@*)     (*@\textcolor{kwQuantum}{\textbf{qv.meas}}@*)  v3, v1, x7, 12
# gate ID 0x68 -> 8 MEASURE; halt asserts
0x30  (*@\textcolor{kwHalt}{\textbf{F0610657}}@*)     (*@\textcolor{kwHalt}{\textbf{qv.resume}}@*) v6, v2, x6, 12
# gate ID 0x78 -> resume marker; halt clears
0x34  0000006F     jal     x0, 0
# self-loop (halt program)
\end{lstlisting}
\vspace{-6pt}
\end{figure}

\subsubsection{\textbf{Pulse Sequence Verification}}
The binary instructions are then loaded into the QVCP, where we monitor the per-qubit issue events at the dispatcher boundary. Listing~\ref{lst:cosim} reports the trace. Note that \texttt{qvsg\_meas} is asserted at \emph{commit time} (cycle~31) rather than at AWG firing time (cycle~106): this conservative early-stall prevents the Ibex core from speculatively committing classical instructions that may depend on the measurement result before the qubit pulse has actually been issued.

The trace records 32 quantum events and 40 qubit firings with zero FIFO errors, in exact agreement with the QASM semantics. The full execution program completes in 177 cycles (1.72\,$\mu$s at the synthesized 100 MHz), of which 124 cycles are waiting for external measurement input. This result confirms that the toolchain---from high-level source to dispatched qubit pulses---is functionally consistent end-to-end.

\begin{figure}[htbp]
\vspace{-8pt}
\centering
\begin{lstlisting}[style=hisepq, language=CoSimTrace,
    caption={Per-qubit dispatch trace},
    label={lst:cosim}]
[cycle=  31] qvsg_meas = 1         
# Ibex halt asserted             
[cycle=  60] AWG: H    on q[0,2,4,6,8,10,12,14]    
[cycle=  83] AWG: CNOT on q[0..15]                  
(CTRL: even, TGT: odd)
[cycle= 106] AWG: MEAS on q[0,2,4,6,8,10,12,14]    
[cycle= 153] external: measure_done = 1             
[cycle= 155] qvsg_meas = 0             
<-- halt cleared
[cycle= 177] AWG: RESUME on q[1,3,5,7,9,11,13,15]   
\end{lstlisting}
\vspace{-8pt}
\end{figure}

\begin{figure*}[ht]
    \centering
    \includegraphics[width=0.8\linewidth]{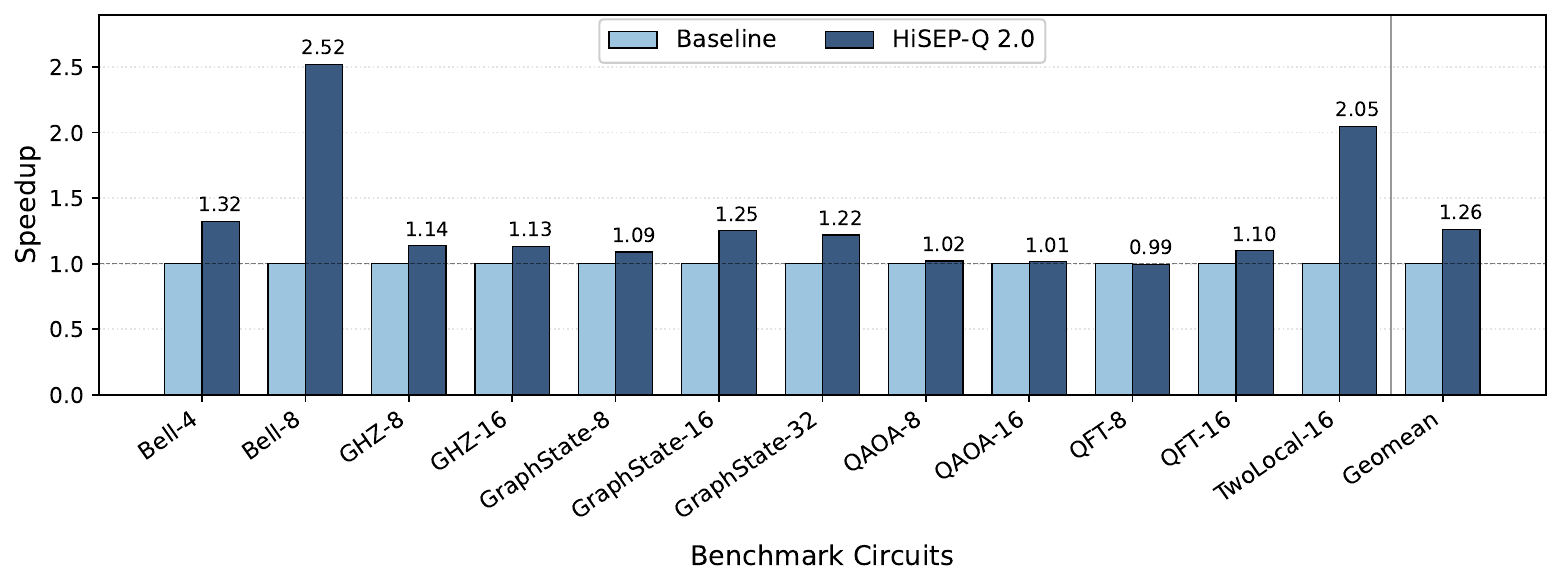}
    \caption{Execution time speedup compared to HiSEP-Q~1.0 (Baseline)}
    \label{fig:speedup}
\end{figure*}

\subsection{Execution Time}

Execution time measures how long the compiled program runs on our \ac{QCP} and is regarded as one of the most important performance metrics. Figure~\ref{fig:speedup} illustrates the execution time speedup of HiSEP-Q~2.0 over HiSEP-Q~1.0 across various MQT Bench workloads~\cite{quetschlich2023mqt}, assuming a 50~ns backend measurement duration. Overall, HiSEP-Q~2.0 achieves a geometric-mean speedup of 1.26$\times$, peaking at 2.52$\times$ for Bell-8. Based on these results, we highlight three key observations:

\textit{1) Scalability with qubit count:} Within each circuit family, speedup grows with the number of qubits: Bell-4$\rightarrow$Bell-8 ($1.32\times\rightarrow2.52\times$), GraphState-8/16/32 ($1.09\times\rightarrow1.25\times$), QFT-8$\rightarrow$QFT-16 ($0.99\times\rightarrow1.10\times$). This demonstrates that larger quantum circuits, in terms of qubit count, can extract greater performance benefits from our vectoring \ac{ISA}. \textit{2) Advantage in parallel operations:}  HiSEP-Q~2.0 delivers substantially larger speedups for algorithms exhibiting high parallelism in two-qubit gate operations. We can observe from the figure that the two largest gains, Bell-8 ($2.52\times$) and TwoLocal-16 ($2.05\times$), both contain layers of independent CX pairs. While baseline \ac{ISA} encodes a single pair per instruction (costing $N$ batches for $N$ parallel CXs), this work packs the entire layer into a single vector CX instruction, explaining this substantial performance leap. \textit{3) Limitations with sequential dependencies:} Circuits with deep sequential dependencies, such as QAOA-8/16 and QFT-8, contain long chains of dependent CX gates that must be executed sequentially. This prevents the \ac{ISA} from packing them into vector instructions. Consequently, HiSEP-Q~2.0 yields marginal to no performance advantage for these specific workloads.

\subsection{Latency and Throughput}

\begin{figure}[htbp]
    \centering
    \begin{subfigure}[b]{0.9\linewidth}
        \centering
        \includegraphics[width=\linewidth]{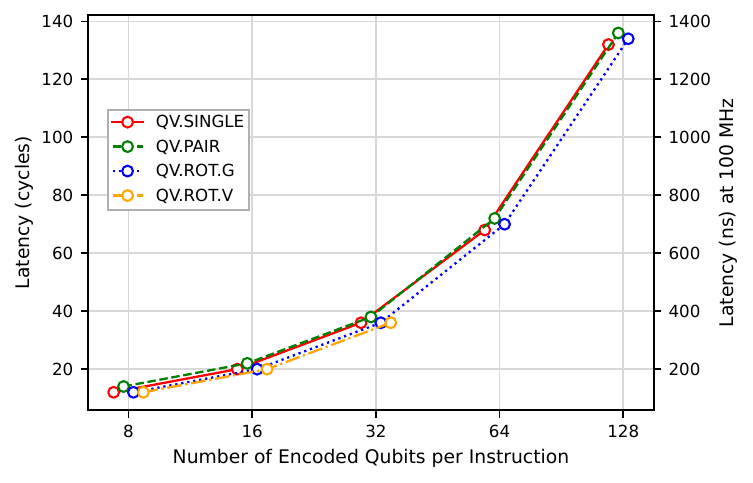}
        \caption{Latency}
        \label{fig:latency}
    \end{subfigure}
    \hfill 
    \begin{subfigure}[b]{0.9\linewidth}
        \centering
        \includegraphics[width=\linewidth]{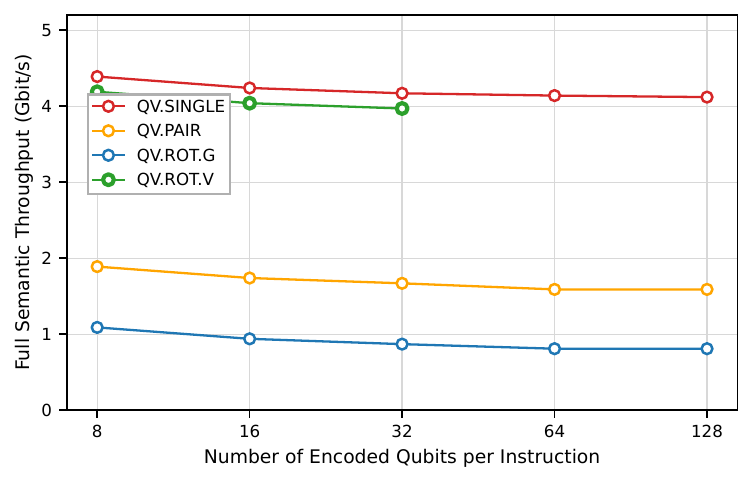}
        \caption{Throughput}
        \label{fig:throughput}
    \end{subfigure}
    \caption{Latency and throughput of the new quantum instructions with different numbers of encoded qubits.}
    \label{fig:performance_metrics}
\end{figure}

Figure~\ref{fig:latency} shows that, at \texttt{SEW=8}, the observed latency of \texttt{QV.SINGLE}, \texttt{QV.ROT.G}, and \texttt{QV.ROT.V} tracks the useful event count almost one-to-one, indicating nearly one emitted event per cycle in steady state. \texttt{QV.PAIR} and \texttt{QV.ROT.G} incur a small excess latency at larger qubit counts due to fixed stream-framing and pipeline-alignment overheads. Grouping factors of \texttt{m4} and \texttt{m8} are illegal for \texttt{QV.ROT.V} and therefore are not shown. In general, the observed latency is roughly linear with the number of encoded qubits; however, there are several configuration instructions before the system triggers the actual quantum event stream. It depends on the grouping factor that assembles multiple physical vector registers into a single logic vector for more qubits, or more cycles to handle the synchronization in the control flow. Regarding the latter case, for example, the halt-resume mechanism of \texttt{QV.SINGLE.MEASURE} adds a distinct control-path delay: after \texttt{measure\_done} is observed, the processor requires 8 cycles (80 $ns$ @100 MHz) to resume normal execution. 

To obtain a practical throughput metric, we do not simply count the full bit width of the exported interface. The fields in the quantum instructions are not equally meaningful across all instructions. We therefore separate the exported data into two categories: per-event dynamic payload and per-instruction semantic metadata. The dynamic payload consists of bits whose values may change from one quantum event to the next, such as qubit indices, paired indices, or rotation angles. The semantic metadata consists of bits that are required to interpret the stream but are typically constant across one instruction instance, such as operation type, destination metadata, and packed fields such as \texttt{block\_imm}. Based on this, the throughput is computed as: 

\begin{equation}
    T = (N_{\mathrm{evt}} B_{\mathrm{dyn}} + B_{\mathrm{meta}}) * F_{qvcp}
\end{equation}

where $N_{\mathrm{evt}}$ is the number of quantum events issued per instruction, $B_{\mathrm{dyn}}$ is the number of semantically meaningful dynamic bits per quantum event, $B_{\mathrm{meta}}$ is the number of configuration bits counted once per instruction rather than once per quantum event, and $F_{qvcp}$ (100 MHz) is the frequency. This calibration avoids overstating throughput by repeatedly counting stream-static metadata, while still preserving the semantic content needed to interpret the quantum event stream.

Figure~\ref{fig:throughput} shows two clear throughput levels. \texttt{QV.SINGLE} remains high and nearly flat because each event carries an 8-bit qubit index together with a 32-bit scalar field. \texttt{QV.ROT.V} exhibits a similar level for its legal configurations (\texttt{mf2}, \texttt{m1}, and \texttt{m2}). By contrast, \texttt{QV.PAIR} and \texttt{QV.ROT.G} are lower because their semantically useful per-event payload is smaller. Their slight decline is caused by fixed framing overheads, since the non-payload cycles are required to start and terminate each quantum stream, to assert the qualifiers, and to align the exported stream with the underlying vector pipeline state. As the number of encoded qubits increases, these constant overhead cycles are amortized over a longer stream, but they remain visible in the measured window and therefore slightly reduce the average throughput.

\subsection{Resource Utilization}
We analyze the resource utilization of HiSEP-Q~2.0 on the ZCU216 FPGA for scalability considerations. To evaluate the cost of each architectural component, we instrument the synthesis flow to produce hierarchical utilization reports and factor the results into the three top-level blocks of the design: the \textit{Ibex} scalar core, the \textit{Vector Engine} (which contains the RVV-based vector core, the dual execution pipelines, and the vector register file), and the \textit{Quantum Dispatcher} that converts vector element streams into per-qubit timed scheduling pulses.

Table~\ref{tab:utilization_32q} summarizes the resource utilization for the 32-qubit configuration. The entire design consumes $14{,}888$ LUTs ($3.50\%$), $15{,}492$ FFs ($1.82\%$), and $10$ DSP blocks ($0.23\%$) on the ZCU216, and requires no BRAMs. The vector register file is mapped entirely to distributed LUT-based memory (LUTRAM), and all internal FIFOs in the quantum dispatcher are small enough to be synthesized from flip-flops rather than dedicated memory blocks. Among the three components, the \textit{Vector Engine} dominates LUT usage ($8{,}540$ LUTs, $2.01\%$) and contains all DSPs used by the multiply pipeline, while the \textit{Quantum Dispatcher} dominates FF usage ($8{,}436$ FFs, $0.99\%$) due to its per-qubit instruction queues (implemented by FIFO). The Ibex core remains lightweight at $3{,}435$ LUTs and $1{,}897$ FFs. Overall, the design occupies less than $4\%$ of the available resources on the device, leaving sufficient space for system integration~\cite{Amr:hpc} and additional quantum control peripherals, like \acp{AWG} and readout processing units~\cite{MLreadout,maurya2023scaling}.

\begin{table}[tb]
\centering
\caption{Resource Utilization of HiSEP-Q 2.0 (32-qubit configuration)}
\label{tab:utilization_32q}
\begin{tabularx}{\linewidth}{l|*{4}{>{\raggedleft\arraybackslash}X|}}
\toprule
\textbf{Component} & \textbf{LUT} & \textbf{LUTRAM} & \textbf{FF} & \textbf{DSP} \\
\midrule
Ibex Core          & 3{,}435 (0.81\%) & --               & 1{,}897 (0.22\%) & 1 (0.02\%) \\
Vector Engine      & 8{,}540 (2.01\%) & 1{,}024 (0.48\%) & 5{,}139 (0.60\%) & 9 (0.21\%) \\
Quantum Dispatcher & 2{,}912 (0.68\%) & --               & 8{,}436 (0.99\%) & --         \\
\midrule
\textbf{Total}     & \textbf{14{,}887 (3.50\%)} & \textbf{1{,}024 (0.48\%)} & \textbf{15{,}472 (1.81\%)} & \textbf{10 (0.23\%)} \\
\bottomrule
\end{tabularx}
\end{table}

\begin{figure}[tb]
    \centering
    \includegraphics[width=1\linewidth]{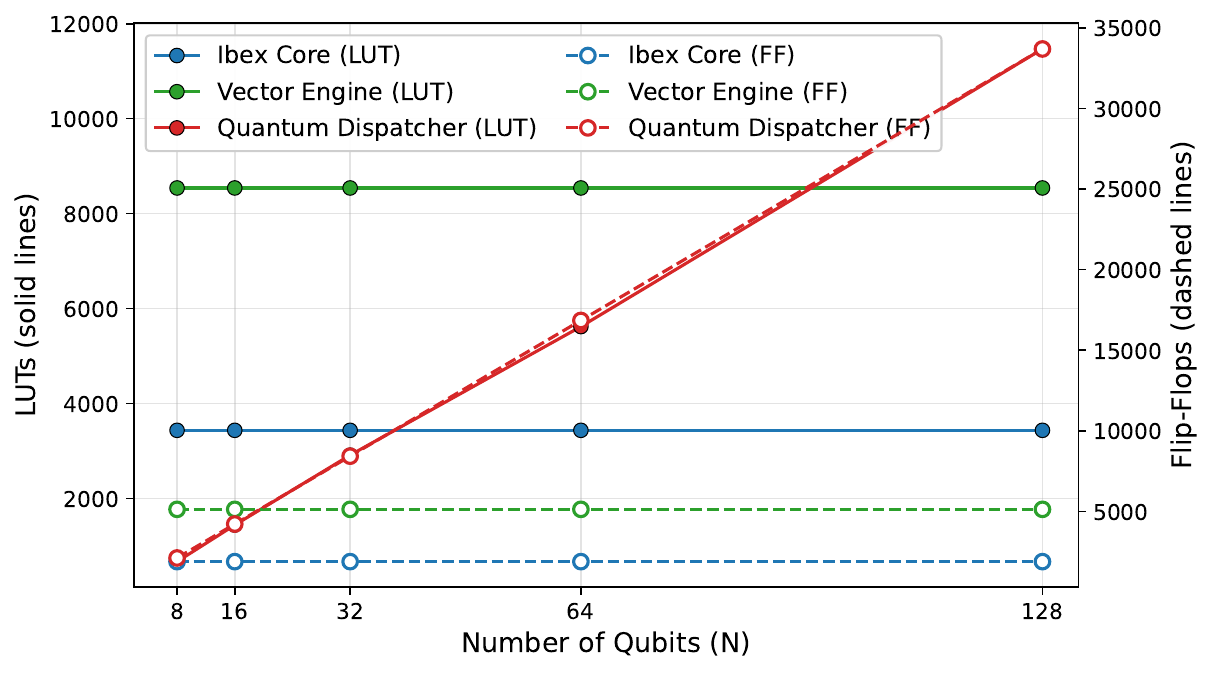}
    \caption{Resource scalability (LUT\&FF) of HiSEP-Q 2.0 versus qubit count $N$.}
    \label{fig:scalability}
\end{figure}

Since the number of controlled qubits ($N$) is the primary scaling parameter for \acp{QCP}, we further investigated how each component of HiSEP-Q~2.0 scales with $N$. We evaluate the performance with $N \in {8, 16, 32, 64, 128}$. Figure~\ref{fig:scalability} plots the measured LUT (solid, left axis) and FF (dashed, right axis) usage of the three components across this range. The \textit{Ibex} core and \textit{Vector Engine} are independent of $N$ by construction: the scalar instruction set, the RVV vector pipelines, and the vector register file do not depend on how many physical qubits are dispatched to downstream pulse generators. As expected, their curves are flat across the entire sweep.

The \textit{Quantum Dispatcher}, in contrast, scales \emph{strictly linearly} in $N$. Each additional qubit instantiates one \texttt{timed\_fifo} block (a per-qubit instruction queue paired with a lightweight timing controller), contributing approximately $90$ LUTs and $263$ flip-flops per qubit. This matches the analytical expectation from the Register-transfer Level (RTL) structure and confirms that the dispatcher imposes no quadratic cost in qubit count. At the largest configuration we tested, $N=128$, the dispatcher consumes $11{,}468$ LUTs and $33{,}684$ FFs, which is a significant but still moderate usage that fits comfortably within the ZCU216's total resources ($425{,}280$ LUTs and $850{,}560$ FFs). This linear scaling behavior further demonstrates that HiSEP-Q~2.0 can be scaled to hundreds of qubits on a single RFSoC device without architectural modifications.

\subsection{Scalability Discussion}
The resource sweep in Figure~\ref{fig:scalability} has already shown that HiSEP-Q~2.0's hardware scales near-linearly with the qubit count $N$ while the scalar/vector front-end remains constant. We now discuss how the same feature holds at the \ac{ISA} level, ensuring that the architecture can extend to future \ac{FTQC} workloads along two further axes: instruction-encoding capacity and vector-length configuration.

\textit{Instruction-encoding capacity.} HiSEP-Q~2.0 reuses the standard RISC-V R-type layout and repurposes \texttt{funct7} as the \texttt{GateID} field, where the 7-bit \texttt{funct7} provides $2^{7}{=}128$ distinct gate encodings per \texttt{funct3} subclass. Reserving all eight \texttt{funct3} values for the quantum extension yields up to $8\times128=1024$ encodable gates. This is more than sufficient to cover any foreseeable logical or physical gate set, including the syndrome-extraction and lattice-surgery primitives required for surface-code \ac{FTQC}.

\textit{Vector-length configuration.} The \texttt{SEW} field determines the addressable qubit space ($2^{\texttt{SEW}}$), so scaling beyond the 128-qubit point reported above only requires reconfiguring \texttt{SEW} at runtime---no ISA change or recompilation of the dispatcher is needed. However, this needs to trade some parallelization for reach: a larger \texttt{SEW} reduces the number of qubits addressed per instruction (Equation~\ref{eq:vl}), but the \texttt{vsetvli}-driven model leaves the choice to the compiler, which can pick the operating point that best matches each algorithmic phase.

Together with the linear hardware scaling, these two ISA-level properties confirm that HiSEP-Q~2.0 is ready for \ac{FTQC}-scale qubit counts without architectural redesign.

%% file: chapters/relatedwork.tex
\section{Related Work}
Numerous prior works have investigated the design of \acp{QCP}. Early efforts~\cite{britt2017isaqpu,fu2018quma,butko2020understanding} focus on quantum \ac{ISA} development, aiming to establish a unified abstraction that describes the interaction between classical and quantum execution. These efforts, however, remain largely at the theoretical level and do not map the proposed \acp{ISA} down to a concrete microarchitecture. Fu et al.~\cite{fu2019eqasm} first propose an executable \ac{ISA} tailored to their respective hardware architectures, and subsequent works~\cite{GuoHiSEP,xu2023qubic,Qtenon} extend this line of research with more scalable and optimized designs. Since each of these designs relies on a custom \ac{ISA}, they require substantial compiler and toolchain development, which limits reusability across platforms and hinders ecosystem-level extensibility. More recently, Liu et al.~\cite{liu2025risc} and Zhao et al.~\cite{HISQ} have started to leverage RISC-V to represent
quantum operations, thereby inheriting its mature open-source
toolchain. However, they rely solely on the RISC-V scalar \ac{ISA} without introducing a quantum-specific extension or leveraging RVV, which makes scaling a fundamental bottleneck in \ac{FTQC}. In contrast, HiSEP-Q~2.0 augments RISC-V with a
compact quantum extension built on top of the RVV framework,
allowing a single vector instruction to dispatch gate operations to an arbitrary number of qubits in parallel while preserving compatibility with the standard RISC-V ecosystem.

%% file: chapters/conclusion.tex
\section{Conclusion}

In this work, we presented HiSEP-Q~2.0, a scalable \ac{QCP} built as an
extension to the RISC-V Vector \ac{ISA}. By mapping qubit addressing
and gate dispatch onto the RVV framework while leaving the scalar
pipeline entirely unmodified, the proposed design inherits the mature
RISC-V toolchain and co-processor ecosystem, eliminating the compiler
burden that has limited prior \ac{QCP} efforts. The extension densely
encodes up to 128 qubit targets in a single instruction through
\ac{LMUL} grouping, and provides dedicated rotation instructions to
support the parameterised gates required by hybrid quantum-classical
algorithms. At the microarchitecture level, we introduce a halt-resume
protocol that tightly coordinates the scalar control core and the
vector pipeline across \ac{MCM} boundaries through a minimal interface.

We verify the design on both the standard RISC-V toolchain, for
binary-compatibility checks, and the ZCU216 FPGA platform, for
on-board performance evaluation. Experimental results demonstrate that
HiSEP-Q~2.0 achieves up to 2.52$\times$ speedup in end-to-end execution
time compared to HiSEP-Q~1.0, while resolving each \ac{MCM}
request within 80\,\text{ns}. The low FPGA resource footprint further
confirms the scalability of the architecture and its suitability for
integration into larger quantum control stacks targeting
\ac{QEC}-scale systems.

%% file: ref.bib
@article{huang2020superconducting,
  title     = {Superconducting quantum computing: a review},
  author    = {Huang, He-Liang and Wu, Dachao and Fan, Daojin and Zhu, Xiaobo},
  journal   = {Science China Information Sciences},
  volume    = {63},
  number    = {8},
  pages     = {Art. no. 180501},
  year      = {2020},
  publisher = {Springer},
  doi       = {10.1007/s11432-020-2881-9}
}

@article{bruzewicz2019trapped,
  title     = {Trapped-ion quantum computing: Progress and challenges},
  author    = {Bruzewicz, Colin D. and Chiaverini, John and McConnell, Robert and Sage, Jeremy M.},
  journal   = {Applied Physics Reviews},
  volume    = {6},
  number    = {2},
  pages     = {Art. no. 021314},
  year      = {2019},
  publisher = {AIP Publishing},
  doi       = {10.1063/1.5088164}
}

@article{bluvstein2026fault,
  title     = {A fault-tolerant neutral-atom architecture for universal quantum computation},
  author    = {Bluvstein, Dolev and Geim, Alexandra A. and Li, Sophie H. and Evered, Simon J. and Bonilla Ataides, J. Pablo and Baranes, Gefen and Gu, Andi and Manovitz, Tom and Xu, Muqing and Kalinowski, Marcin and others},
  journal   = {Nature},
  volume    = {649},
  number    = {8095},
  pages     = {39--46},
  year      = {2026},
  publisher = {Nature Publishing Group UK London}
}

@article{abughanem2025ibm,
   title={IBM quantum computers: evolution, performance, and future directions},
   volume={81},
   ISSN={1573-0484},
   url={http://dx.doi.org/10.1007/s11227-025-07047-7},
   DOI={10.1007/s11227-025-07047-7},
   number={5},
   journal={The Journal of Supercomputing},
   publisher={Springer Science and Business Media LLC},
   author={AbuGhanem, Muhammad},
   year={2025},
   month=Apr }

@misc{liu2025risc,
  title         = {{RISC-Q}: A Generator for Real-Time Quantum Control System-on-Chips Compatible with {RISC-V}},
  author        = {Liu, Junyi and Lee, Yi and Deng, Haowei and Clayton, Connor and Yang, Gengzhi and Wu, Xiaodi},
  year          = {2025},
  eprint        = {2505.14902},
  archivePrefix = {arXiv},
  primaryClass  = {quant-ph}
}

@article{stefanazzi2022qick,
  title     = {The {QICK} ({Quantum Instrumentation Control Kit}): Readout and control for qubits and detectors},
  author    = {Stefanazzi, Leandro and Treptow, Kenneth and Wilcer, Neal and Stoughton, Chris and Bradford, Collin and Uemura, Sho and Zorzetti, Silvia and Montella, Salvatore and Cancelo, Gustavo and Sussman, Sara and others},
  journal   = {Review of Scientific Instruments},
  volume    = {93},
  number    = {4},
  pages     = {Art. no. 044709},
  year      = {2022},
  publisher = {AIP Publishing},
  doi       = {10.1063/5.0076249}
}

@inproceedings{GuoHiSEP,
  author    = {Guo, Xiaorang and Qin, Kun and Schulz, Martin},
  booktitle = {2023 IEEE 41st International Conference on Computer Design (ICCD)},
  title     = {{HiSEP-Q}: A Highly Scalable and Efficient Quantum Control Processor for Superconducting Qubits},
  year      = {2023},
  pages     = {86--93},
  doi       = {10.1109/ICCD58817.2023.00023}
}

@misc{AMDrfsoc,
  author       = {{Advanced Micro Devices, Inc.}},
  title        = {{AMD Zynq UltraScale+ RFSoCs}: The Industry's Only Single-Chip Adaptable Radio Platform},
  year         = {2025},
  howpublished = {\url{https://www.amd.com/en/products/adaptive-socs-and-fpgas/soc/zynq-ultrascale-plus-rfsoc.html}},
  note         = {Accessed: 2025}
}

@misc{xu2023qubic,
  title         = {{Qubic 2.0}: An extensible open-source qubit control system capable of mid-circuit measurement and feed-forward},
  author        = {Xu, Yilun and Huang, Gang and Fruitwala, Neelay and Rajagopala, Abhi and Naik, Ravi K. and Nowrouzi, Kasra and Santiago, David I. and Siddiqi, Irfan},
  year          = {2023},
  eprint        = {2309.10333},
  archivePrefix = {arXiv},
  primaryClass  = {quant-ph}
}

@article{wintersperger2023neutral,
  title     = {Neutral atom quantum computing hardware: performance and end-user perspective},
  author    = {Wintersperger, Karen and Dommert, Florian and Ehmer, Thomas and Hoursanov, Andrey and Klepsch, Johannes and Mauerer, Wolfgang and Reuber, Georg and Strohm, Thomas and Yin, Ming and Luber, Sebastian},
  journal   = {EPJ Quantum Technology},
  volume    = {10},
  number    = {1},
  pages     = {Art. no. 32},
  year      = {2023},
  publisher = {Springer},
  doi       = {10.1140/epjqt/s40507-023-00190-1}
}

@inproceedings{Qtenon,
  author    = {Tao, Chenning and Lu, Liqiang and Zheng, Size and Chang, Li-Wen and Shen, Minghua and Zhang, Hanyu and Liu, Fangxin and Zhou, Kaiwen and Yin, Jianwei},
  title     = {{Qtenon}: Towards Low-Latency Architecture Integration for Accelerating Hybrid Quantum-Classical Computing},
  booktitle = {Proceedings of the 52nd Annual International Symposium on Computer Architecture (ISCA '25)},
  year      = {2025},
  publisher = {Association for Computing Machinery},
  address   = {New York, NY, USA},
  pages     = {299--312},
  isbn      = {9798400712616},
  doi       = {10.1145/3695053.3731087}
}

@inproceedings{HISQ,
  author    = {Zhao, Yilun and Zhao, Kangding and Zhou, Peng and Liu, Dingdong and Luo, Tingyu and Zheng, Yuzhen and Luo, Peng and Hu, Shun and Lin, Jin and Guo, Cheng and Han, Yinhe and Wang, Ying and Deng, Mingtang and Wu, Junjie and Fu, X.},
  title     = {{Distributed-HISQ}: A Distributed Quantum Control Architecture},
  booktitle = {Proceedings of the 58th IEEE/ACM International Symposium on Microarchitecture (MICRO '25)},
  year      = {2025},
  publisher = {Association for Computing Machinery},
  address   = {New York, NY, USA},
  pages     = {564--578},
  isbn      = {9798400715730},
  doi       = {10.1145/3725843.3756048}
}

@article{fu2018quma,
  author  = {Fu, Xubin and Rol, Menno A. and Bultink, C. C. and van Someren, Jeroen and Khammassi, Nader and Ashraf, Imran and Vermeulen, R. F. L. and De Sterke, J. C. and Vlothuizen, W. J. and Schouten, R. N. and Almud{\'e}ver, C. G. and Bertels, Koen and DiCarlo, Leonardo},
  title   = {A Microarchitecture for a Superconducting Quantum Processor},
  journal = {IEEE Micro},
  year    = {2018},
  volume  = {38},
  number  = {3},
  pages   = {40--47},
  doi     = {10.1109/MM.2018.032271060}
}

@inproceedings{fu2019eqasm,
  author    = {Fu, Xubin and Riesebos, Lisanne and Rol, Menno A. and van Straten, Jeroen and van Someren, Jeroen and Khammassi, Nader and Ashraf, Imran and Vermeulen, R. F. L. and Newsum, Vincent and Loh, K. K. L. and De Sterke, J. C. and Vlothuizen, W. J. and Schouten, R. N. and Almud{\'e}ver, Carmen G. and DiCarlo, Leonardo and Bertels, Koen},
  title     = {{eQASM}: An Executable Quantum Instruction Set Architecture},
  booktitle = {2019 IEEE International Symposium on High Performance Computer Architecture (HPCA)},
  year      = {2019},
  pages     = {224--237},
  doi       = {10.1109/HPCA.2019.00040}
}

@inproceedings{britt2017isaqpu,
  author    = {Britt, Keith A. and Humble, Travis S.},
  title     = {Instruction Set Architectures for Quantum Processing Units},
  booktitle = {High Performance Computing. ISC High Performance 2017 International Workshops},
  series    = {Lecture Notes in Computer Science},
  volume    = {10524},
  year      = {2017},
  pages     = {98--105},
  publisher = {Springer},
  doi       = {10.1007/978-3-319-67630-2_8}
}

@article{britt2017hpcqpu,
  author  = {Britt, Keith A. and Humble, Travis S.},
  title   = {High-Performance Computing With Quantum Processing Units},
  journal = {ACM Journal on Emerging Technologies in Computing Systems},
  year    = {2017},
  volume  = {13},
  number  = {3},
  pages   = {Art. no. 39},
  doi     = {10.1145/3007651}
}

@manual{rvv_spec,
  author       = {{RISC-V International}},
  title        = {The {RISC-V} ``V'' Vector Extension, Version 1.0},
  organization = {RISC-V International},
  year         = {2021},
  url          = {https://github.com/riscv/riscv-v-spec},
  note         = {Ratified specification}
}

@inproceedings{platzer2021vicuna,
  author    = {Platzer, Michael and Puschner, Peter},
  title     = {{Vicuna}: A Timing-Predictable {RISC-V} Vector Coprocessor for Scalable Parallel Computation},
  booktitle = {33rd Euromicro Conference on Real-Time Systems (ECRTS 2021)},
  series    = {Leibniz International Proceedings in Informatics (LIPIcs)},
  volume    = {196},
  year      = {2021},
  pages     = {1:1--1:18},
  doi       = {10.4230/LIPIcs.ECRTS.2021.1}
}

@manual{ibexDocs,
  author       = {{lowRISC contributors}},
  title        = {{Ibex}: An Embedded 32-Bit {RISC-V} {CPU} Core},
  organization = {lowRISC CIC},
  year         = {2026},
  url          = {https://ibex-core.readthedocs.io/en/latest/},
  note         = {Accessed: Apr. 7, 2026}
}

@misc{microsoft_quantum_stack,
  author       = {{Microsoft}},
  title        = {The Quantum Computing Stack},
  year         = {2024},
  howpublished = {\url{https://quantum.microsoft.com/en-us/insights/education/concepts/quantum-computing-stack}},
  note         = {Accessed: 2024-05-22}
}

@article{peruzzo2014variational,
  title     = {A variational eigenvalue solver on a photonic chip},
  author    = {Peruzzo, Alberto and McClean, Jarrod and Shadbolt, Peter and Yung, Man-Hong and Zhou, Xiao-Qi and Love, Peter J. and Aspuru-Guzik, Al{\'a}n and O'Brien, Jeremy L.},
  journal   = {Nature Communications},
  volume    = {5},
  number    = {1},
  pages     = {Art. no. 4213},
  year      = {2014},
  publisher = {Springer Nature},
  doi       = {10.1038/ncomms5213}
}

@misc{farhi2014quantum,
  title         = {A Quantum Approximate Optimization Algorithm},
  author        = {Farhi, Edward and Goldstone, Jeffrey and Gutmann, Sam},
  year          = {2014},
  eprint        = {1411.4028},
  archivePrefix = {arXiv},
  primaryClass  = {quant-ph}
}

@article{xu2021qubic,
  title     = {{QubiC}: An open-source {FPGA}-based control and measurement system for superconducting quantum information processors},
  author    = {Xu, Yilun and Huang, Gang and Balewski, Jan and Naik, Ravi and Morvan, Alexis and Mitchell, Bradley and Nowrouzi, Kasra and Santiago, David I. and Siddiqi, Irfan},
  journal   = {IEEE Transactions on Quantum Engineering},
  volume    = {2},
  pages     = {1--11},
  year      = {2021},
  publisher = {IEEE},
  doi       = {10.1109/TQE.2021.3116540}
}

@techreport{xif_spec,
  title       = {{CORE-V eXtension Interface (XIF)} Specification},
  author      = {{OpenHW Group}},
  institution = {OpenHW Group},
  year        = {2023},
  note        = {Available at \url{https://docs.openhwgroup.org/projects/openhw-group-core-v-xif/}}
}

@inproceedings{schlaegl2024rvvvp,
  author    = {Schlaegl, Manfred and Stockinger, Moritz and Gro{\ss}e, Daniel},
  title     = {A {RISC-V} ``V'' {VP}: Unlocking Vector Processing for Evaluation at the System Level},
  booktitle = {2024 Design, Automation \& Test in Europe Conference (DATE)},
  year      = {2024},
  url       = {https://ics.jku.at/files/2024DATE_RISCV-VP-plusplus_RVV.pdf}
}

@manual{llvm_rvv,
  author       = {{LLVM Project}},
  title        = {{LLVM}: {RISC-V} Vector Extension},
  organization = {LLVM Documentation},
  year         = {2025},
  url          = {https://llvm.org/docs/RISCV/RISCVVectorExtension.html},
  note         = {Accessed: 2026-04-15}
}

@inproceedings{Amr:hpc,
  author    = {Elsharkawy, Amr and Guo, Xiaorang and Schulz, Martin},
  booktitle = {2024 IEEE International Conference on Quantum Computing and Engineering (QCE)},
  title     = {Integration of Quantum Accelerators into {HPC}: Toward a Unified Quantum Platform},
  year      = {2024},
  volume    = {01},
  pages     = {774--783},
  doi       = {10.1109/QCE60285.2024.00097}
}

@inproceedings{MLreadout,
  author    = {Guo, Xiaorang and Bunarjyan, Tigran and Liu, Dai and Lienhard, Benjamin and Schulz, Martin},
  booktitle = {2025 62nd ACM/IEEE Design Automation Conference (DAC)},
  title     = {{KLiNQ}: Knowledge Distillation-Assisted Lightweight Neural Network for Qubit Readout on {FPGA}},
  year      = {2025},
  pages     = {1--7},
  doi       = {10.1109/DAC63849.2025.11132854}
}

@inproceedings{maurya2023scaling,
  title     = {Scaling qubit readout with hardware efficient machine learning architectures},
  author    = {Maurya, Satvik and Mude, Chaithanya Naik and Oliver, William D. and Lienhard, Benjamin and Tannu, Swamit},
  booktitle = {Proceedings of the 50th Annual International Symposium on Computer Architecture (ISCA '23)},
  pages     = {1--13},
  year      = {2023},
  publisher = {Association for Computing Machinery},
  doi       = {10.1145/3579371.3589042}
}

@inproceedings{butko2020understanding,
  title        = {Understanding quantum control processor capabilities and limitations through circuit characterization},
  author       = {Butko, Anastasiia and Michelogiannakis, George and Williams, Samuel and Iancu, Costin and Donofrio, David and Shalf, John and Carter, Jonathan and Siddiqi, Irfan},
  booktitle    = {2020 International Conference on Rebooting Computing (ICRC)},
  pages        = {66--75},
  year         = {2020},
  organization = {IEEE},
  doi          = {10.1109/ICRC2020.2020.00011}
}

@article{quetschlich2023mqt,
  title     = {{MQT Bench}: Benchmarking software and design automation tools for quantum computing},
  author    = {Quetschlich, Nils and Burgholzer, Lukas and Wille, Robert},
  journal   = {Quantum},
  volume    = {7},
  pages     = {Art. no. 1062},
  year      = {2023},
  publisher = {Verein zur F{\"o}rderung des Open Access Publizierens in den Quantenwissenschaften},
  doi       = {10.22331/q-2023-07-20-1062}
}
